\documentclass[letter]{aa} 
\usepackage{graphicx}
\usepackage{txfonts}
\usepackage{natbib}
\bibpunct{(}{)}{;}{a}{}{,}
\newcommand{\doceCO}{\mbox{$^{12}$CO}}
\newcommand{\treceCO}{\mbox{$^{13}$CO}}
\newcommand{\jdu}{\mbox{$J$=2$-$1}}
\newcommand{\juc}{\mbox{$J$=1$-$0}}

\newcommand{\kms}{\mbox{km\,s$^{-1}$}}

\newcommand{\minky}{\mbox{M~1--92}}
\newcommand{\etacar}{\mbox{$\eta$~Car}}
\newcommand{\homunculus}{\mbox{\em ``Homunculus''}}
\newcommand{\Jybeam}{\mbox{Jy\,beam$^{-1}$}}
\newcommand{\msun}{\mbox{$M_{\mbox{\sun}}$}}
\newcommand{\lsun}{\mbox{$L_{\mbox{\sun}}$}}
\newcommand{\my}{\mbox{$M_{\mbox{\sun}}$\,yr$^{-1}$}}

\newcommand{\TB}{\mbox{$T_{\mathrm B}$}}
\newcommand{\Tkin}{\mbox{$t_{\mathrm{kin}}$}}

\begin{document}
   \title{Minkowski's Footprint revisited}

   \subtitle{Planetary Nebula formation from a single sudden event~?~\thanks{Based on 
             observations carried out with the IRAM Plateau de Bure Interferometer. 
             IRAM is supported by INSU/CNRS (France), MPG (Germany) and IGN (Spain).}}

   \author{
          J. Alcolea\inst{1}
          \and
          R. Neri\inst{2}
          \and
          V. Bujarrabal\inst{3}
          }

   \offprints{J. Alcolea, \email{j.alcolea[at]oan.es} }

   \institute{Observatorio Astron\'omico Nacional (OAN-IGN), 
              Calle Alfonso XII 3, E-28014 Madrid, Spain. 
              \and
              Institut de Radio Astronomie Millim\'etrique (IRAM),
              300 Rue de la Piscine, F-38406 St.-Martin d'H\`eres, France 
              \and
              Observatorio Astron\'omico Nacional (OAN-IGN),
              Apartado 112, E-28803 Alcal\'a de Henares, Spain 
              }

\date{Received December 18, 2006; accepted January 12, 2007}

  \abstract
   {\minky\ can be considered an archetype of bipolar pre-planetary nebulae.
It shows a clear axial symmetry, along with the kinematics and momentum excess characteristic
of this class of envelopes around post-AGB stars.}
   {By taking advantage of the new extended configuration of the IRAM Plateau de Bure
interferometer, we wanted to study the morphology and velocity field of 
the molecular gas better in this nebula, particularly in its central part.}
   {We performed sub-arcsecond resolution interferometric observations of the \jdu\ rotational line 
of \treceCO\ in \minky.}
   {We found that the equatorial component is a thin flat disk, which expands radially
with a velocity proportional to the distance to the center. The kinetic age of this
equatorial flow is very similar to that of the two lobes. The small widths and
velocity dispersion in the gas forming the lobe walls confirm that the acceleration 
responsible for the nebular shape could not last more than 100--120\,yr.}
   {The present kinematics of the molecular gas can be explained
as the result of a single brief acceleration event, after which the nebula reached an expansion velocity 
field with axial symmetry. 
In view of the similarity to other objects, we speculate
on the possibility that the whole nebula was formed as a 
result of a magneto-rotational explosion
in a common-envelope system.}

   \maketitle
%

\section{Introduction}

PN~\minky, also known as Minkowski's Footprint, is an
O-rich pre-planetary nebula (pPN) originally discovered by \citet{minkowski},
which has been deeply observed in the optical, NIR, and radio wavelengths.
The central star has a spectrum compatible with $T_{\mathrm{eff}}$ = 6500~K and 
$\log g =$ 0.5, although at short wavelengths it shows an excess that
can be fitted with the presence of a secondary star of $T_{\mathrm{eff}}$ = 18\,000~K and 
$\log g =$ 5.0 \citep{arrieta}. The distance to \minky\ has been
estimated as 2.5~kpc, adopting a nominal luminosity of 10$^4$~\lsun\ for 
a post-AGB star \citep{cohenkuhi}. The nebula has a size 
of 11\arcsec\  in the axial direction and 6\arcsec\ in the transversal one.
It consists of a two-lobe reflection nebula, divided by a dense equatorial 
component. The two lobes define a clear axis of symmetry, oriented at 
a position angle (P.A.) of 311\degr\ (measured from north to east).
This axis is inclined in respect to the plane of the sky by
about 35\degr, with the northwest lobe pointing to us. Optical spectroscopy shows 
the presence of wide absorption components, revealing the existence of (bipolar) 
flows of ionized gas close to the star, with expansion velocities up to 750~\kms\ 
\citep{arrieta}. Line emission in H$\alpha$ and in atomic forbidden lines 
(\ion{O}{i}, \ion{O}{iii}, \ion{N}{ii}, \ion{S}{ii}) is 
detected from the middle of the two lobes \citep{bujhst}, 
probably tracing the location of shocks propagating along these jets 
\citep[see also][]{solf}. \citet{bujhst} estimated
that the mass of the ionized gas in the nebula is only about 10$^{-3}$\,\msun, and 
\citet{arrieta} concluded that the present ionized mass loss is very low, 3\,10$^{-13}$\,\my.

  \begin{figure*}
   \centering
   \includegraphics{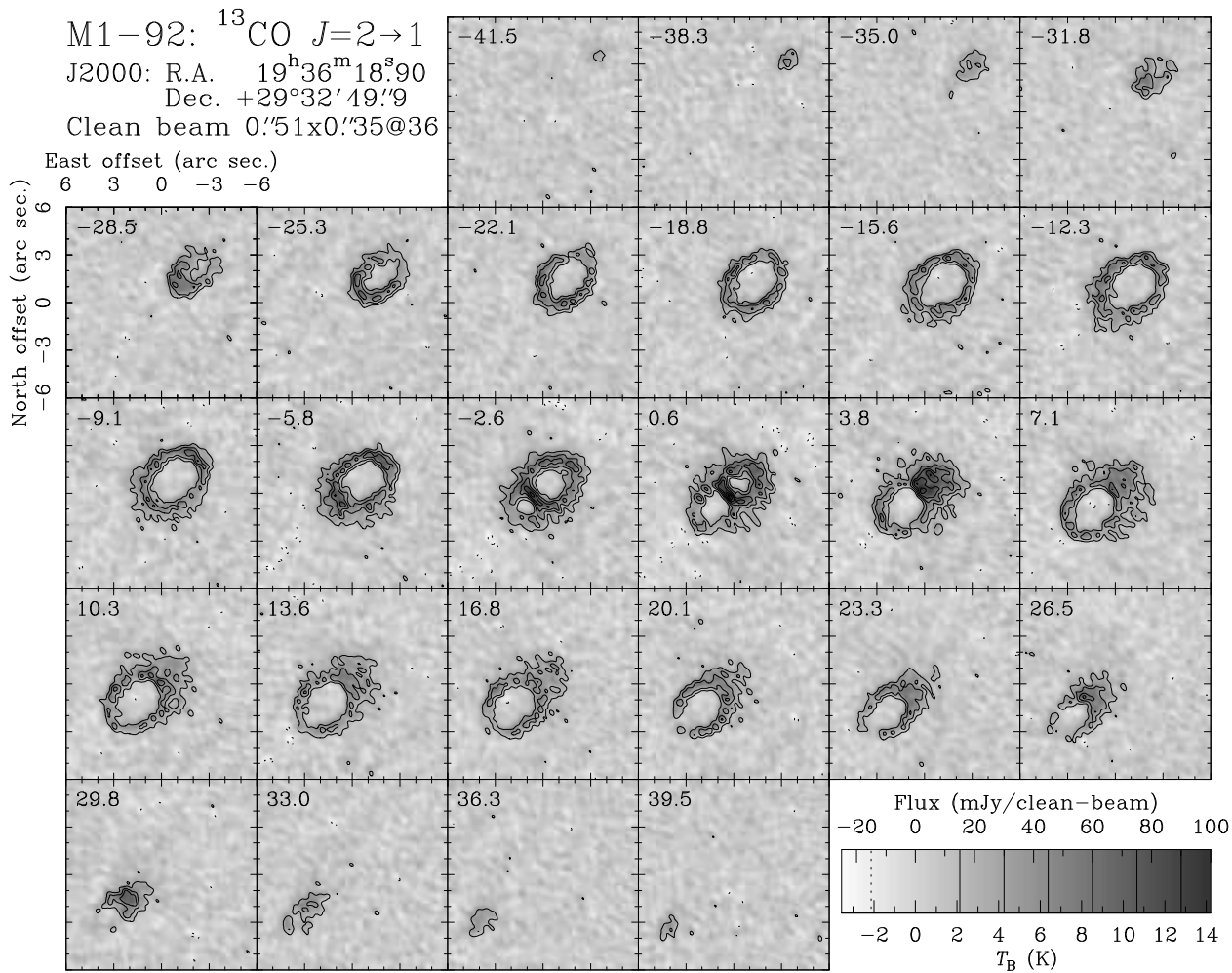}
   \caption{The 3.25\,\kms\ resolution channel maps 
     of the intensity of \treceCO\ \jdu\ in \minky\ from the new+old Plateau
            de Bure data (see text). The LSR velocity in \kms\ for the center of each channel
            is indicated in the upper left corner. Drawn contours are $-$15 (dashed), and 15 to 90 by
            10\,m\Jybeam. The assumed coordinates for the reference 
            position are indicated in the top left corner of the plot.}
          
              \label{maps}%
    \end{figure*}

Because of the early stage of the central star in its post-AGB evolution, almost all the 
nebular material is still in the form of molecular gas. 
This component has been studied in detail by \citet{bujpdb} by means of interferometric 
observations of the \juc\ and \jdu\ rotational lines of \treceCO, obtaining a resolution
of $\sim$1\farcs0 at 220\,GHz (in the \jdu\ line). The molecular gas is located in the
equatorial disk and in the walls of the two lobes, which are practically empty. From
those maps, the total mass estimated for the molecular component is 0.9\,\msun, 0.2\,\msun\ of which 
is located in the disk and 0.35\,\msun\ in the walls of each lobe. The excitation was found
to be very low across the nebula, from 10 to 15\,K, and the densities ranging from 5\,10$^{4}$ to
3\,10$^{5}$\,cm$^{-3}$, the warmer and densest parts located close to the equator. 
The velocity field of the gas was also investigated using a spectral resolution of 3.25\,\kms.  
These authors found that the kinematics consists in a low radial expansion component of 8\,\kms, 
which dominates in the equatorial region, plus an axial component that increases linearly 
with the distance to the equatorial plane, reaching an expansion velocity of 70\,\kms at
the two nebular tips. Such a Hubble-like velocity law was interpreted as the result of 
a sudden interaction that lasted much less than the kinetic age of the axial outflow
($\sim$\,1000\,yr). In addition, an axial momentum of 22\,\msun\,\kms\ was measured in the nebula. This value 
is much larger that what can be provided by radiation pressure on dust grains on the expected 
acceleration lime (about 100\,yr or less). In fact these two properties, the presence of 
Hubble-like axial acceleration, and the momentum excess are very common among pPNe
\citep{bujalas}.

  \begin{figure*}
   \centering
   \includegraphics{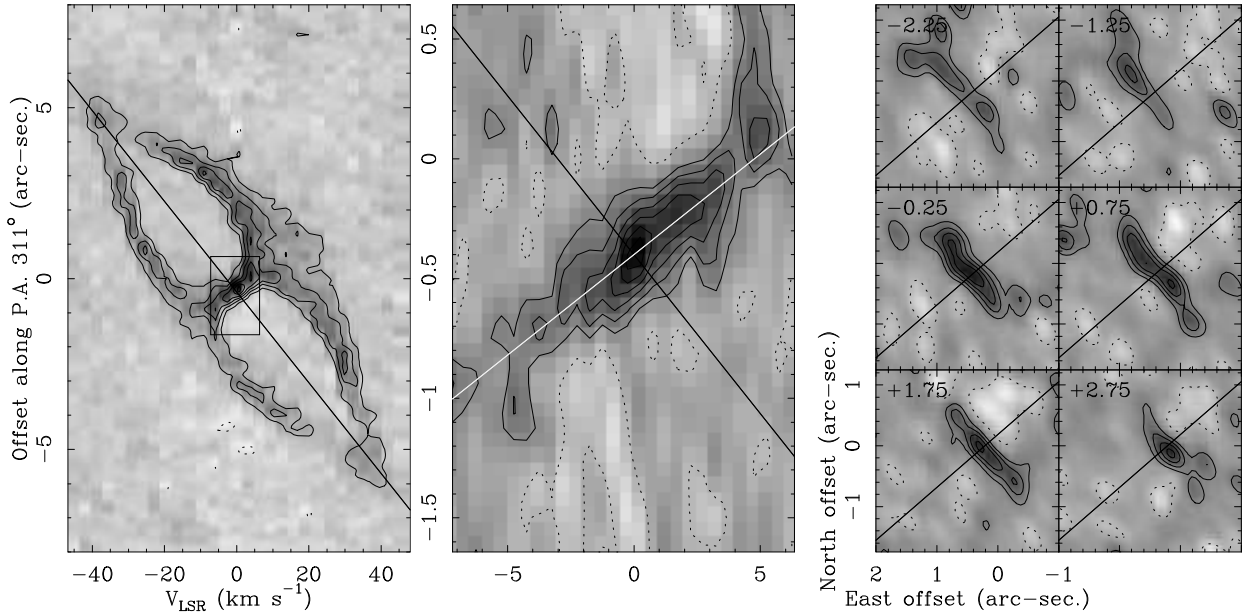}
   \caption{{\bf Left:} Position vs. velocity diagram for a cut along the symmetry axis of the nebula
            (P.A. 311\degr). Contours are as in Fig~\ref{maps}. The straight line shows the velocity
            gradient found along this direction. {\bf Center:} The same for the central region of the nebula
            (the rectangle shown on the left) from the maps obtained only with 6Aq data and for a spectral 
            resolution of 0.5\,\kms. Contours are as before. The white/black straight lines  show the
            gradient of the equatorial/extended components of the nebula. {\bf Right:} Selected channel 
            velocity maps of the center of the nebula, obtained just from the 6Aq data with 0.5\,\kms\ 
            resolution. Contours are as before. The straight line delineate the axial direction.
            (See also Fig.\,\ref{maps}.)} 
              \label{cortes}%
    \end{figure*}

In spite of all these observational efforts, very little is known about the mechanism that
accelerated and stretched the nebula along its symmetry axis, as well as about what
is powering the present bipolar outflow seen in ionized gas. (Note that the 
jet seen in the optical does not have enough energy or momentum to explain the present kinematics, 
so
we may be in the presence of two different processes.) This is partially due to the
limited spatial resolution of the molecular line observation, which cannot reveal
the structure of the lobe walls or the shape and size of the equatorial component.
Making use of the new, improved resolution of the Plateau de Bure interferometer, in this
letter we report on new sub-arcsecond resolution observations of \minky\ in the \treceCO\ 
\jdu\ line.
 

\section{New Plateau de Bure observations}

These new observations of \minky\ were conducted with the new
extended configurations 6Aq \& 6Bq of the IRAM Plateau de Bure
mm-wave interferometer. The
observations were performed on January 24
and February 26, 2006 (configurations A and B, respectively),
under excellent to good  weather conditions. Using the 1.3~mm receiver, we 
observed the \jdu\ rotational line of \treceCO, adopting  a rest 
frequency of 230\,399~MHz. The \juc\ lines of \treceCO\ and
C\element[][18]{O}\ were simultaneously observed with the 3~mm receiver, but 
their results will be discussed elsewhere. The correlator was set up to
cover the whole \jdu\ line of \treceCO\ with an original spectral resolution 
of 156.25~kHz, i.e. 0.20~\kms\ at 230.4~GHz; however, to ease the comparison
with previous maps of the nebula, we re-sampled the channels to 
the same velocity resolution used before. To improve the reliability 
of the phase calibration, water vapor corrections based on the results of 
the 22~GHz radiometers installed on the antennas were applied. Final phase 
calibration was performed from the observations of the quasars J1923+210 and J2023+336
(inter-spaced every 20~min.), which were also used as secondary amplitude 
calibrators. Observations of the radio-sources 3C~273, 3C~345, and MWC~349 
were performed to fix the absolute flux scale and to calibrate the gain across the band.

After the data set was 
calibrated, maps were produced and CLEANed. The maps from the new data showed 
a significant fraction of lost flux, $\sim$\,50\% on average, when compared with previous 
observations. These losses are smaller, only $\sim$\,30\%, for velocities that are both 
extreme and close to the systemic one. In contrast, for intermediate expansion 
velocities, 5--20~\kms, the missing flux fraction is as high as $\sim$\,2/3.
In order to recover this lost emission and produce maps with both high-contrast and
spatial resolution, we merged our new data with those from our 1995--1997
observations \citep{bujpdb}, using robust weighting. The resulting 
clean-beam has an HPBW of 0\farcs51$\times$0\farcs35, with the major axis oriented 
at P.A. = 36\degr. This clean beam was later used in the CLEANing
procedure following the Clark method. Comparing the recovered flux in this data set with that
from our previous observations both with Pico de Veleta and Plateau de Bure, we
conclude that all spectra are compatible within the uncertainties and that no flux
is lost in the maps. The final r.m.s. for the adopted spectral resolution of
3.25\,\kms\ is 4.5\,m\Jybeam, and the conversion from these flux units into 
Rayleigh-Jeans equivalent brightness temperature \TB\ is 142\,K per \Jybeam.
In addition we produced maps using just the visibilities from the most extended configuration (6Aq) 
and a spectral resolution of 0.5\,\kms, to better investigate the structure and velocities in the
equatorial component. The resulting clean-beam has an HPBW of 0\farcs46$\times$0\farcs26 at 
P.A. of 25\degr, the final r.m.s. is 15\,m\Jybeam, and the conversion from these 
flux units into \TB\ is 208~K per \Jybeam.

\section{Observational results}

The maps are shown in Fig.\,\ref{maps} and the corresponding position vs. velocity diagram along the symmetry 
axis is displayed in Fig.\,\ref{cortes} (left panel). The 
observations clearly show the structure of the nebula, with its
two empty lobes divided by the equatorial disk, and the axial Hubble-like velocity field. The lobe walls are 
very thin, with a de-convolved FWHP of 0\farcs25 (9\,10$^{15}$\,cm). The inner limits of these walls are much 
better defined than the outer boundary, where some weak diffuse emission is detected, 
especially at low velocities. 
The velocity dispersion in the walls must also be very small, since we can hardly detect it with a 3.25\,\kms\ 
resolution. At the nebular tips, these widths and velocity dispersion are larger, about 0\farcs9 and 
5\,\kms, respectively. All these figures are typically 10--15\% of the distance to the center  and of the 
expansion velocity at the corresponding points. This puts a strong upper limit on the duration of the 
forces responsible for
accelerating the nebula along the symmetry axis; we confirm that the process could not last more than about 
one tenth of the kinetic age of the flow: i.e. $\la$\,100--120\,yr (see later). 

At the center of the nebula we find that the equatorial structure is not compatible with our previous model.
We do not see the signatures of a disk expanding at a constant radial velocity 
\citep[see e.g.][]{chiu2006pigru}.
 This is much better displayed with
the results from the 6Aq configuration (Fig.\,\ref{cortes}, center and right panels). 
In the axial velocity cut, 
we find that this structure also shows a constant velocity gradient, but in the opposite sense. In the maps, 
we see that the emission at each velocity traces a strip parallel to the symmetry axis. All these features can
only be fitted with a disk in which the radial expansion is not constant but proportional to the distance to
the center: {\em i.e. we also find a Hubble-like velocity field in the equatorial plane, a totally unexpected result!} 
Note that the maximum in the maps arises from the center of this disk and at the systemic velocity,
in total disagreement with the expectations for an equatorial undisturbed remnant of an AGB envelope.
From the extent of the emission at the systemic velocity, we derive that the diameter of the disk is 1\farcs4 
(5\,10$^{16}$\,cm) and the height at half power is $\la$\,0\farcs2 (7\,10$^{15}$\,cm). 
At the resolution achieved in the maps, we do not find a central cavity or signatures of gas rotation.
 
For the axial and equatorial outflows, we find a projected velocity gradient of 7.6 and 12\,\kms\ per 
arc-second, respectively. Assuming the distance and inclination with respect to the plane of the sky given 
in Sect.~1, 
these figures translate into kinetic ages, \Tkin, of 1060 and 1370\,yr, respectively. However, since the two 
flows are perpendicular to each other, these kinetic ages have an opposite dependence with the adopted 
inclination angle. In particular, changing this angle from 35\degr\ to 38.5\degr\ (i.e. only by 10\% 
and within the uncertainties in our previous modeling), both
ages become the same, 1200\,yr. This inclination is still compatible the one derived by 
\citet{solf} of 33\degr$\pm$5\degr, and the resulting coeval scenario is much more plausible than having one 
ejection 
along the equatorial plane 300\,yr before the the axial flow is launched. If 
this hypothesis is correct, the whole 
CO nebula would share the same Hubble-like velocity field: the velocity would just be
proportional to the distance, $ \vec{V}=\vec{R}/\Tkin$, and we can simply use the velocity along the line of sight 
to derive the gas distribution in this direction. In this case the position vs. velocity 
digram shown in 
Fig.\,\ref{cortes} would represent the spatial structure of \minky, if we forget our finite resolution in 
velocity and space, by simply translating the angular distance and velocities into length units using the 
distance to the target and the derived velocity gradient. Note that even the location of the weak diffuse 
emission seen along the equatorial plane at velocities around 10--15\,\kms\ (Fig.\,\ref{cortes} left) seems 
consistent with the halo detected in the maps in Fig.\,\ref{maps}.

\section{The origin of the pPN~\minky}

The origin of bipolar outflows in pPNe is not yet understood. Current scenarios involve the presence of 
accretion on the central star or on the companion in the case of binary systems through a rotating-disk, 
magneto-centrifugal launching, and a variety of collimating agents \citep{balickfrank}. The question of whether 
close binary systems are mandatory for producing axis-symmetric nebulae is also a long-standing issue 
\citep{bujasoker}. Of course explaining the formation of a nebula involving coeval
axial and equatorial Hubble-like flows becomes even more difficult. 

However, \minky\ is not the only object in 
which complex structures resulting from a single sudden event have been found. The brightest luminous
blue variable, \etacar, is surrounded by a cloud of gas and dust, whose central and densest part is 
known as the \homunculus. 
This nebula, with an estimated mass $\sim$\,1\msun, consists of a two-lobe structure divided 
by a thin equatorial torus. \citet{morse} performed a proper motion study in this nebula by comparing HST 
images taken at three different epochs and concluded 
that the \homunculus\ also shows a single Hubble law and was 
formed in less that 10~yr during the great eruption in the 1840's. These measurements are only done in the 
plane of the sky (neither the position nor 
the velocities are derived along the line of sight, as in our case), so they do not depend on
the orientation of the nebula or on the distance. Although \etacar\ and the central system in \minky\ are 
very different, the striking similarity found in the kinematics and shape of these two nebulae strongly 
suggests that they may share a common origin. \citet{matt2006} have developed a model  for \etacar\
in which the differential rotation between the stellar nucleus and its envelope can drive a magneto-rotational
explosion.
In this model, as a result of the sudden release of the magnetic energy when it overcomes the weight of the
upper layers, a bipolar jet is launched but an equatorial flow is driven too. These authors 
suggest that this mechanism would also explain the shapes of bipolar pPNe like IRAS~17106--3046 
\citep{matt2004}. 
Recently \citet{nordhaus2006} conclude that for this mechanism to operate, how the differential
rotation is sustained against the increasing magnetic force can be explained 
by the angular momentum transferred to the 
stellar envelope by a low-mass companion during a common envelope phase. If this is also the case for \minky,
we should conclude that this pPNe is largely formed by gas directly expelled from the stellar envelope and
not a former AGB circumstellar envelope accelerated by the interaction with some post-AGB 
flow. Otherwise it would be very difficult to explain how the present gas could show such an accurate  
ballistic movement.

Although \minky\ is so far the only case in which we can argue that the central disk was formed at the same 
time as the lobes,  there are other pPNe in which radial acceleration has been found along the equator in 
addition to the typical bipolar flows. In CRL~2688 \citet{cox2688} found in the CO maps up to 7 different 
bipolar outflows, three along the axis of symmetry (D to F in their naming) and four more 
(A to C and G) that could be running in different directions in the equatorial plane. In all these jets the 
velocity increases with the 
distance to the center. The authors do not comment on whether a Hubble law also holds here, but from Fig.~3 in 
that paper we conclude that a constant velocity gradient is consistent with the data. However, it is very 
difficult to see whether these equatorial flows have similar kinetic ages 
(so as to  compare them with those for the axial
jets). For the axial flows, this 
Hubble-law velocity field has been recently confirmed by \citet{ueta2006}, by means of proper motion
measurements using HST NICMOS images. 
Another interesting case is CRL~618. Here \citet{carmen618} observed \doceCO, and when fitting
their results with a numerical model, they found that an increasing velocity
was also necessary for the central dense core. This model neither assumed an exact Hubble law field nor 
constrained this kinematics
to a thin equatorial disk, but note that the fitting was not very precise in this region because of the limited
spatial resolution of the observations. (In these two cases, the interaction of the post-AGB
flows with the old AGB shell is clearly detected.) Finally, Frosty~Leo presents a dense equatorial 
disk/torus that 
expands at higher velocities than those typical in AGB envelopes. \citet{frostyleo} conclude that such a
high velocity was probably due to interactions between the AGB and post-AGB flows in the a plane perpendicular 
to the axial direction. However, in view of our current finding, we may also speculate with the 
hypothesis of an equatorial 
acceleration in this source, too. Note, however, that \minky\ is one the very few pPN in which
the molecular gas has been studied at such high spatial and spectral resolution. Therefore it would 
very premature to conclude whether the mechanism proposed here for explaining the kinematics of
\minky\ should also apply to other pPNe. We must wait until high-resolution images of the molecular gas in
more sources are available, but we think that this possibility should be explored in the context of 
understanding the issues of pPNe and PNe formation and shaping.

\begin{acknowledgements}
      J.A.and V.B. acknowledge their partial support from the 
      \emph{Spanish Ministry of Education and Science} under projects AYA2003-7584 and ESP2003-04957. 
     Data reduction and plots have been performed using the Gildas software package.
\end{acknowledgements}

\bibliographystyle{aa}
\bibliography{minky}
\end{document}